\documentclass[12pt]{article}

\usepackage{cite}
\usepackage{graphicx}
\usepackage{pict2e}
\usepackage[all]{xy}

\author{Yu.~M.~Zinoviev
       \thanks{E-mail address: Yurii.Zinoviev@ihep.ru} \\[0.5cm]
        {\it Institute for High Energy Physics} \\
        {\it of National Research Center "Kurchatov Institute"} \\
        {\it Protvino, Moscow Region, 142280, Russia}}

\title{On massive higher spins and gravity.\\
 IV. Arbitrary spin}

\date{}

\begin{document}

\maketitle

\begin{abstract}
In this paper, we investigate gravitational interactions of massive
fields with arbitrary integer and half-integer spin, trying to
construct a vertex that contains both standard minimal and
non-minimal interaction terms necessary to make the vertex gauge
invariant. We propose an ansatz for these non-minimal terms and show
that it leads to a unique solution that correctly reproduces our
previous results for spins 5/2, 3 and 7/2, including all possible
partially massless limits.
\end{abstract}

\thispagestyle{empty}
\newpage
\setcounter{page}{1}

\section{Introduction}
 
In recent papers \cite{Zin25a,Zin25b,Zin25c}, we have investigated
gravitational interactions for massive fields with spin 5/2, 3 and
7/2, including all possible partially massless limits. The main goal
was to construct vertices that contain standard minimal interactions
and have as low number of derivatives as possible. Here, we
summarize the main results of our work.
\begin{itemize}
\item There exist non-singular massless limits in anti de Sitter space
$\Lambda < 0$. It means, in-particular, that the maximal number of
derivatives is the same as in the massless case.
\item There exist non-singular flat limits for non-zero mass.
\item In all three cases partially massless fields with depth $k=2$
have minimal gravitational interactions.
\item Partially massless fields for $s=3,\frac{7}{2}$ and depth $k=3$
as well as spin 7/2 fields living on the boundary of unitary forbidden
region do not have minimal interactions.
\end{itemize}

In this work we present a generalization of the previous results for
the massive fields with arbitrary integer and half-integer spin. As
before, we use the frame-like gauge invariant formalism for
massive higher spin fields \cite{Zin08b,PV10,KhZ19} in its multispinor
version \cite{KhZ19} and use the same notation and conventions. In
Section 2 we provide all the necessary kinematical information, and in
Section 3, we discuss gravitational interactions. Our main
proposal is a general ansatz for non-minimal interactions, which is
just the sum of non-minimal terms for all helicities $h \ge 2$ with
the coefficients depending on both mass and cosmological constant. We
show that this ansatz gives a unique solution that correctly
reproduces all previous results.

\section{Kinematics}

\subsection{General analysis}

The main objects in the frame-like gauge invariant multispinor
formalism \cite{KhZ19} are one-form fields
$\Phi^{\alpha(k),\dot\alpha(l)}$ with some numbers of completely
symmetric dotted and undotted spinor indices $\alpha,\dot\alpha=1,2$.
Note that the fields with $k+l$ even correspond to bosons and those
with $k+l$ odd correspond to fermions. The Lorentz invariance
strongly restricts the possible structure of the gauge
transformations, and the most general ansatz is given by\footnote{For
coordinate free description of (anti) de Sitter space we use
background frame $e^{\alpha\dot\alpha}$ and Lorentz covariant
derivative $D$ such that
$$
D \wedge e^{\alpha\dot\alpha} = 0, \qquad
D \wedge D \zeta^\alpha = 2\Lambda E^\alpha{}_\beta \zeta^\beta,
\qquad e^{\alpha\dot\alpha} \wedge e^{\beta\dot\beta}
= \epsilon^{\dot\alpha\dot\beta} E^{\alpha\beta} +
\epsilon^{\alpha\beta} E^{\dot\alpha\dot\beta}.
$$}:
\begin{eqnarray}
\delta \Phi^{\alpha(k)\dot\alpha(l)} &=& 
D \zeta^{\alpha(k)\dot\alpha(l)} + e_\beta{}^{\dot\alpha}
\zeta^{\alpha(k)\beta\dot\alpha(l-1)} + a_{k,l} e_{\beta\dot\beta}
\zeta^{\alpha(k)\beta\dot\alpha(l)\dot\beta} \nonumber \\
 && + b_{k,l} e^\alpha{}_{\dot\beta} 
\zeta^{\alpha(k-1)\dot\alpha(l)\dot\beta} + c_{k,l}
e^{\alpha\dot\alpha} \zeta^{\alpha(k-1)\dot\alpha(l-1)},
\end{eqnarray}
Moreover, this in turn determines the structure of the gauge invariant
two-forms (curvatures)\footnote{The curvatures with $k=0$ or $l=0$
require introduction of the so-called Stueckelberg zero-forms, but we
will not need such fields in what follows.}:
\begin{eqnarray}
{\cal F}^{\alpha(k)\dot\alpha(l)} &=& D \Phi^{\alpha(k)\dot\alpha(l)}
+ e_\beta{}^{\dot\alpha} \Phi^{\alpha(k)\beta\dot\alpha(l-1)} +
a_{k,l} e_{\beta\dot\beta} \Phi^{\alpha(k)\beta\dot\alpha(l)\dot\beta}
\nonumber \\
 && + b_{k,l} e^\alpha{}_{\dot\beta} 
\Phi^{\alpha(k-1)\dot\alpha(l)\dot\beta} + c_{k,l}
e^{\alpha\dot\alpha} \Phi^{\alpha(k-1)\dot\alpha(l-1)}.
\end{eqnarray}
Let us stress that such construction connects only nearest neighbours
but does not mix bosonic and fermionic fields. 
$$
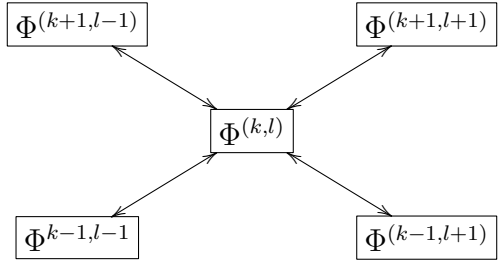
\xymatrix{
*+[F]{\Phi^{(k+1,l-1)}} \ar@{<->}[dr] & & 
*+[F]{\Phi^{(k+1,l+1)}} \ar@{<->}[dl] \\
 & *+[F]{\Phi^{(k,l)}} & \\
*+[F]{\Phi^{k-1,l-1}} \ar@{<->}[ur] & & 
*+[F]{\Phi^{(k-1,l+1)}} \ar@{<->}[ul] }
$$
Calculating variations of the curvatures under the gauge
transformations we obtain
\begin{eqnarray}
\delta {\cal F}^{\alpha(k)\dot\alpha(l)} &=& 
[ 2\Lambda - l b_{k+1,l-1} + (l+2)b_{k,l} 
+ (l+2)a_{k,l}c_{k+1,l+1} - la_{k-1,l-1}c_{k,l} ]
E^\alpha{}_\beta \zeta^{\alpha(k-1)\beta\dot\alpha(l)} \nonumber \\
 && + [ 2\Lambda + (k+2) b_{k+1,l-1} - k b_{k,l} 
+ (k+2)a_{k,l}c_{k+1,l+1} - k a_{k-1,l-1}c_{k,l} ]
E^{\dot\alpha}{}_{\dot\beta} \zeta^{\alpha(k)\dot\alpha(l-1)\dot\beta}
\nonumber \\
 && + [ (l+2)a_{k,l} - l a_{k+1,l-1}]
 E_{\beta(2)} \zeta^{\alpha(k)\beta(2)\dot\alpha(l)} \nonumber \\
 && + 2[ (k+2) c_{k+1,l-1} - kc_{k,l} ] 
 E^{\dot\alpha(2)} \zeta^{\alpha(k)\dot\alpha(l-2)} \nonumber \\
 && + [ (k+2)a_{k,l}b_{k+1,l+1} - k a_{k-1,l+1}b_{k,l}]
 E_{\dot\beta(2)} \zeta^{\alpha(k)\dot\alpha(l)\dot\beta(2)} \nonumber
\\
 && + 2[ (l+2)c_{k-1,l+1}b_{k,l} - lc_{k,l}b_{k-1,l-1}]
 E^{\alpha(2)} \zeta^{\alpha(k-2)\dot\alpha(l)},
\end{eqnarray}
thus the gauge invariance requires that the coefficients satisfy
\begin{eqnarray}
0 &=& 2\Lambda - l b_{k+1,l-1} + (l+2)b_{k,l} 
+ (l+2)a_{k,l}c_{k+1,l+1} - la_{k-1,l-1}c_{k,l}, \nonumber \\
0 &=& 2\Lambda + (k+2) b_{k+1,l-1} - k b_{k,l} 
+ (k+2)a_{k,l}c_{k+1,l+1} - k a_{k-1,l-1}c_{k,l}, \nonumber \\
0 &=& (l+2)a_{k,l} - l a_{k+1,l-1}, \nonumber \\
0 &=& (k+2) c_{k+1,l-1} - kc_{k,l}, \label{req} \\
0 &=& (k+2)a_{k,l}b_{k+1,l+1} - k a_{k-1,l+1}b_{k,l}, \nonumber \\
0 &=& (l+2)c_{k-1,l+1}b_{k,l} - lc_{k,l}b_{k-1,l-1}. \nonumber
\end{eqnarray}
To construct a gauge invariant description of massive higher spins, we
need to find solutions of these general relations that are consistent
with the free Lagrangians and their symmetries. However, the free
Lagrangians of bosons and fermions differ significantly, so we must
consider these two cases separately.

\subsection{Integer spin}

To construct a gauge invariant Lagrangian for massive boson with mass
$M$ and spin $s$ we need pairs of physical and auxiliary one-forms
($f^{\alpha(m)\dot\alpha(m)}$, $\Omega^{\alpha(m+1)\dot\alpha(m-1)} +
h.c$), $1 \le m \le s-1$, one-form $A$ and Stueckelberg zero-forms
($B^{\alpha(2)} + h.c.$, $\pi^{\alpha\dot\alpha}$, $\varphi$). This
Lagrangian has the form:
\begin{eqnarray}
\frac{1}{i} {\cal L}_0 &=& \sum_{m=1}^{s-1} (-1)^{m+1} 
[ (m+1) \Omega^{\alpha(m)\beta\dot\alpha(m-1)} E_\beta{}^\gamma
\Omega_{\alpha(m)\gamma\dot\alpha(m-1)} \nonumber \\
 && \qquad \qquad  - (m-1)
\Omega^{\alpha(m+1)\dot\alpha(m-2)\dot\beta} 
E_{\dot\beta}{}^{\dot\gamma} 
\Omega_{\alpha(m+1)\dot\alpha(m-2)\dot\gamma} \nonumber \\
 && \qquad \qquad +  2 \Omega^{\alpha(m)\beta\dot\alpha(m-1)}
e_\beta{}^{\dot\beta} D f_{\alpha(m)\dot\alpha(m-1)\dot\beta}] +
h.c. \nonumber \\
 && + [4E B_{\alpha(2)} B^{\alpha(2)} + 2 E_{\alpha(2)} B^{\alpha(2)}
D A + h.c.] - 6 E \pi_{\alpha\dot\alpha} \pi^{\alpha\dot\alpha} - 12
E_{\alpha\dot\alpha} \pi^{\alpha\dot\alpha} D \varphi \nonumber \\
 && + \sum_{m=2}^{s-1} (-1)^m a_m [ E_{\beta(2)}
\Omega^{\alpha(m-1)\beta(2)\dot\alpha(m-1)} 
f_{\alpha(m-1)\dot\alpha(m-1)} \nonumber \\
 && \qquad \qquad + \frac{(m-1)}{(m+1)} E_{\beta(2)}
f^{\alpha(m-2)\beta(2)\dot\alpha(m)}
\Omega_{\alpha(m-2)\dot\alpha(m)} + h.c. \nonumber \\
 && + a_1 [ \Omega^{\alpha(2)} E_{\alpha(2)} A - 2 B^{\beta\alpha}
E_\beta{}^{\dot\beta} f_{\alpha\beta} + h.c. ] + a_0
E_{\alpha\dot\alpha} \pi^{\alpha\dot\alpha} A \nonumber \\
 && + \sum_{m=1}^{s-1} (-1)^{m+1} b_m 
[ f^{\alpha(m-1)\beta\dot\alpha(m)} E_\beta{}^\gamma 
f_{\alpha(m-1)\gamma\dot\alpha(m)} + h.c. ] \nonumber \\
 && + \frac{a_1a_0}{2} E_{\alpha\dot\alpha}
f^{\alpha\dot\alpha} \varphi + 3a_1{}^2 E \varphi^2.
\end{eqnarray}
Here
\begin{eqnarray}
b_m &=& \frac{2s(s+1)}{m(m+1)(m+2)}[M^2 - s(s-1)\Lambda], \nonumber \\
a_m{}^2 &=& \frac{4(s-m)(s+m+1)}{m(m-1)} 
[ M^2 - (s-m-1)(s+m)\Lambda] \qquad m \ge 2, \nonumber \\
a_1{}^2 &=& 2(s-1)(s+2) [ M^2 - (s-2)(s+1)\Lambda], \\
a_0{}^2 &=& 24s(s+1)[M^2 - s(s-1)\Lambda]. \nonumber 
\end{eqnarray}
This Lagrangian is invariant under the following gauge transformations
\begin{eqnarray}
\delta f^{\alpha(m)\dot\alpha(m)} &=&  D \xi^{\alpha(m)\dot\alpha(m)}
+ e_\beta{}^{\dot\alpha} \eta^{\alpha(m)\beta\dot\alpha(m-1)} +
e^\alpha{}_{\dot\beta} \eta^{\alpha(m-1)\dot\alpha(m)\dot\beta}
\nonumber \\
 && + \frac{m}{2(m+2)}a_{m+1} e_{\beta\dot\beta} 
\xi^{\alpha(m)\beta\dot\alpha(m)\dot\beta} + \frac{a_m}{2m(m+1)}
e^{\alpha\dot\alpha} \xi^{\alpha(m-1)\dot\alpha(m-1)}, \nonumber \\
\delta \Omega^{\alpha(m+1)\dot\alpha(m-1)} &=& D
\eta^{\alpha(m+1)\dot\alpha(m-1)} + \frac{a_{m+1}}{2} 
e_{\beta\dot\beta} \eta^{\alpha(m+1)\beta\dot\alpha(m-1)\dot\beta}
 + e_\beta{}^{\dot\alpha} \eta^{\alpha(m+1)\beta\dot\alpha(m-2)}
\nonumber \\
 && + \frac{b_m}{2(m+1)}  e^\alpha{}_{\dot\beta}
\xi^{\alpha(m)\dot\alpha(m-1)\dot\beta}
+ \frac{a_m}{2(m+1)(m+2)} e^{\alpha\dot\alpha}
\eta^{\alpha(m)\dot\alpha(m-2)},  \\
\delta f^{\alpha\dot\alpha} &=& D \xi^{\alpha\dot\alpha} 
+ e_\beta{}^{\dot\alpha} \eta^{\alpha\beta} + e^\alpha{}_{\dot\beta}
\eta^{\dot\alpha\dot\beta} + \frac{a_2}{6} e_{\beta\dot\beta}
\xi^{\alpha\beta\dot\alpha\dot\beta} - \frac{a_1}{4}
e^{\alpha\dot\alpha} \xi, \nonumber \\
\delta B^{\alpha(2)} &=& \frac{a_1}{2} \eta^{\alpha(2)}, \qquad
\delta A = D \xi - \frac{a_1}{2} e_{\alpha\dot\alpha}
\xi^{\alpha\dot\alpha}, \nonumber \\
\delta \pi^{\alpha\dot\alpha} &=& - \frac{a_1a_0}{24}
\xi^{\alpha\dot\alpha}, \qquad \delta \varphi = \frac{a_0}{12}.
\nonumber 
\end{eqnarray}
An explicit dependence on the mass $M$ and the cosmological constant 
$\Lambda$ shows that, in de Sitter space with $\Lambda > 0$ there
exists a so-called unitary forbidden region $M^2 < s(s-1)\Lambda$.
Within this region, there are the so-called partially massless
fields, characterized by a depth of partially masslessness $k$ ($k=1$
corresponding to massless case). Namely, whenever one of the
coefficients $a_m=0$, the Lagrangian decomposes into two separate
parts, one containing higher helicities describing partially massless
field. For illustration, let us use spin 4.
\begin{itemize}
\item $M^2 = 12\Lambda$: the boundary of unitary forbidden region and
partially massless field with depth $k=4$ and helicities
$(\pm 4, \pm 3, \pm 2, \pm 1)$;
\item $M^2 = 10\Lambda$: partially massless field with depth $k=3$ and
helicities $(\pm 4, \pm 3, \pm 2)$;
\item $M^2 = 6\Lambda$: partially massless field with depth $k=2$ and
helicities $(\pm 4,\pm 3)$.
\end{itemize}

Now to find a solution of the recurrent relations (\ref{req}) we use
\begin{equation}
k = m + n, \qquad l = m - n,
\end{equation}
so that $n=0,1$ correspond to physical and auxiliary fields, while 
$n \ge 2$ describe so-called extra fields. Then the complete set of
gauge transformations $1 \le m \le s-1$, $1 \le n \le m-1$ looks like:
\begin{eqnarray}
\delta \Omega^{\alpha(m+n)\dot\alpha(m-n)} &=& D 
\eta^{\alpha(m+n)\dot\alpha(m-n)} + e_\beta{}^{\dot\alpha}
\eta^{\alpha(m+n)\beta\dot\alpha(m-n-1)} + a_{m,n}
e_{\beta\dot\beta} \zeta^{\alpha(m+n)\beta\dot\alpha(m-n)\dot\beta}
\nonumber \\
 && + b_{m,n} e^\alpha{}_{\dot\beta} 
\eta^{\alpha(m+n-1)\dot\alpha(m-n)\dot\beta} + c_{m,n}
e^{\alpha\dot\alpha} \eta^{\alpha(m+n-1)\dot\alpha(m-n-1)}
\end{eqnarray}
and similarly for the gauge invariant curvatures. Here
\begin{eqnarray}
b_{m,n} &=& \frac{(s+n)(s-n+1)}{(m+n)(m+n+1)(m-n+1)(m-n+2)}
[ M^2 - (s-n)(s+n-1)\Lambda], \nonumber \\
a_{m,n} &=& \frac{m(m+1)}{2(m-n+2)(m-n+1)}a_{m+1}, \\
c_{m,n} &=& \frac{1}{2(m+n)(m+n+1)}a_m. \nonumber
\end{eqnarray}
These results show, in-particular, that in the partially massless
limit not only Lagrangian fields, but the extra fields also decouple.

\subsection{Half-integer spin}

To describe massive fermion with mass $M$  and spin $s+1/2$ we need
one-forms $\Phi^{\alpha(m+1)\dot\alpha(m)}+h.c.$, 
$0 \le m \le s-1$ and zero-form $\phi^\alpha + h.c.$
The free Lagrangian has the form
\begin{eqnarray}
{\cal L}_0 &=& \sum_{m=0}^{s-1} (-1)^{m+1} 
\Phi_{\alpha(m)\beta\dot\alpha(m)} e^\beta{}_{\dot\beta}
D \Phi^{\alpha(m)\dot\alpha(m)\dot\beta} - \phi_\alpha
E^\alpha{}_{\dot\alpha} D \phi^{\dot\alpha} \nonumber \\
 && + \sum_{m=1}^{s-1} (-1)^{m+1} c_m [ E^{\beta(2)}
\Phi_{\alpha(m-1)\beta(2)\dot\alpha(m)} 
\Phi^{\alpha(m-1)\dot\alpha(m)}] + c_0 \Phi_\alpha
E^\alpha{}_{\dot\alpha} \phi^{\dot\alpha} + h.c. \nonumber \\
 && + \sum_{m=0}^{s-1} (-)^{m+1} \frac{d_m}{2} [ (m+2) 
\Phi_{\alpha(m)\beta\dot\alpha(m)} E^\beta{}_\gamma
\Phi^{\alpha(m)\gamma\dot\alpha(m)} \nonumber \\
 && \qquad \qquad - m \Phi_{\alpha(m+1)\dot\alpha(m-1)\dot\beta}
E^{\dot\beta}{}_{\dot\gamma} 
\Phi^{\alpha(m+1)\dot\alpha(m-1)\dot\gamma}] + d_0 E
\phi_\alpha \phi^\alpha + h.c. 
\end{eqnarray}
Here
\begin{eqnarray}
d_m &=& \frac{(s+1)}{(m+1)(m+2)}\sqrt{M^2 - s^2\Lambda}, \nonumber \\
c_m{}^2 &=& \frac{(s-m)(s+m+2)}{(m+1)^2} 
[ M^2 - (s^2-(m+1)^2)\Lambda], \\
c_0{}^2 &=& 2s(s+2) [ M^2 - (s^2-1)\Lambda]. \nonumber 
\end{eqnarray}
This Lagrangian is invariant under the following gauge transformations
\begin{eqnarray}
\delta \Phi^{\alpha(m+1)\dot\alpha(m)} &=& D 
\zeta^{\alpha(m+1)\dot\alpha(m)} + e_\beta{}^{\dot\alpha}
\zeta^{\alpha(m+1)\beta\dot\alpha(m-1)} + d_m
e^\alpha{}_{\dot\beta} \zeta^{\alpha(m)\dot\alpha(m)\dot\beta}
\nonumber \\
 && + c_{m+1} e_{\beta\dot\beta} 
\zeta^{\alpha(m+1)\beta\dot\alpha(m)\dot\beta} + 
\frac{c_m}{m(m+2)} e^{\alpha\dot\alpha}
\zeta^{\alpha(m)\dot\alpha(m-1)}, \\
\delta \phi^\alpha &=& c_0 \zeta^\alpha. \nonumber
\end{eqnarray}
As in the bosonic case, there is a unitary forbidden region
$M^2 < s^2\Lambda$ and several partially massless fields
living inside (the main difference being that no single partially
massless case coincides with a boundary). Let us use spin 9/2 as an
illustration:
\begin{itemize}
\item $M^2 = 16\Lambda$: the boundary of unitary forbidden region;
\item $M^2 = 15\Lambda$: partially massless field with depth $k=4$ and
helicities $(\pm 9/2, \pm 7/2, \pm 5/2, \pm 3/2)$;
\item $M^2 = 12\Lambda$: partially massless field with depth $k=3$ and
helicities $(\pm 9/2, \pm 7/2, \pm 5/2)$;
\item $M^2 = 7\Lambda$: partially massless field with depth $k=2$ and
helicities $(\pm 9/2, \pm 7/2)$;
\end{itemize}

Now to find a solution of general recurrent relations (\ref{req})
consistent with the free Lagrangian and its symmetries we use
\begin{equation}
k = m + n + 1, \qquad l = m - n,
\end{equation}
so that again $n=0$ corresponds to the physical fields. In this case
the general gauge transformations $1 \le m \le s-1$, $1 \le n \le m-1$
look like:
\begin{eqnarray}
\delta \Phi^{\alpha(m+n+1)\dot\alpha(m-n)} &=& D
\zeta^{\alpha(m+n+1)\dot\alpha(m-n)} + e_\beta{}^{\dot\alpha}
\zeta^{\alpha(m+n+1)\beta\dot\alpha(m-n-1)} 
 + b_{m,n} e^\alpha{}_{\dot\beta} 
\zeta^{\alpha(m+n)\dot\alpha(m-n)\dot\beta} \nonumber \\
 && + c_{m,n} e^{\alpha\dot\alpha} 
\zeta^{\alpha(m+n)\dot\alpha(m-n-1)} + a_{m,n}
e_{\beta\dot\beta} \zeta^{\alpha(m+n+1)\beta\dot\alpha(m-n)\dot\beta}.
\end{eqnarray}
We obtain:
\begin{eqnarray}
a_{m,n} &=& \frac{(m+1)(m+2)}{(m-n+2)(m-n+1)}c_{m+1}, \nonumber \\
c_{m,n} &=& \frac{(m+1)}{m(m+n+1)(m+n+2)}c_m, \\
b_{m,n} &=& \frac{(s+n+1)(s-n+1)}{(m+n+1)(m+n+2)(m-n+1)(m-n+2)}
 [ M^2 - (s^2 - n^2)\Lambda],   \nonumber \\
b_{m,0} &=& d_m. \nonumber
\end{eqnarray}

\section{Gravitational interactions}

\subsection{Massless case}

As is well known \cite{FV87,FV87a,Vas11,KhZ20a}, massless bosons with
spin $s$ can have standard minimal gravitational interactions in anti
de Sitter space $\Lambda = - \lambda^2 < 0$, provided the vertex also
contains non-minimal terms of the form:
\begin{equation}
{\cal L}_{non-min} = \sum_{n=0}^{s-2} \frac{\kappa_n}{\lambda^{2n}}
R_{\dot\alpha\dot\beta} 
\Omega^{\alpha(s-1+n)\dot\gamma(s-2-n)\dot\alpha}
\Omega_{\alpha(s-1+n)\dot\gamma(s-2-n)}{}^{\dot\beta} + h.c.
\end{equation}
In this, a flat limit is possible only after the rescaling of coupling
constant leaving the non-minimal terms with the highest number of
derivatives\footnote{As for the number of derivatives,
recall that the auxiliary field $\Omega^{\alpha(m+1)\dot\alpha(m-1)}$
on-shell is equivalent to the first derivative of the physical field
$f^{\alpha(m)\dot\alpha(m)}$, while the extra field
$\Omega^{\alpha(m+n)\dot\alpha(m-n)}$ is equivalent to the n-th
derivative.} $N=2s-2$
\begin{equation}
{\cal L}_1 \sim R_{\dot\alpha\dot\beta}
\Omega^{\alpha(2s-3)\dot\alpha} \Omega_{\alpha(2s-3)}{}^{\dot\beta} +
h.c. 
\end{equation}
Similarly, for the massless spin $s+1/2$ fermions  the non-minimal
interactions look like:
\begin{equation}
{\cal L}_{non-min} = \sum_{n=0}^{s-2} \frac{\kappa_n}{\lambda^{2n+1}}
R_{\dot\alpha\dot\beta} 
\Phi^{\alpha(s+n)\dot\gamma(s-2-n)\dot\alpha}
\Phi_{\alpha(s+n)\dot\gamma(s-2-n)}{}^{\dot\beta} + h.c.
\end{equation}
while a flat limit (after appropriate rescaling) gives
\begin{equation}
{\cal L}_1 \sim R_{\dot\alpha\dot\beta}
\Phi^{\alpha(2s-2)\dot\alpha} \Phi_{\alpha(2s-2)}{}^{\dot\beta} +
h.c. 
\end{equation}

\subsection{General analysis}

Let us consider the following general ansatz for the non-minimal
interactions:
\begin{equation}
{\cal L}_{non-min} = R_{\dot\alpha\dot\beta} \sum_{k,l} \kappa_{k,l}
\Phi^{\alpha(k)\dot\gamma(l-1)\dot\alpha}
\Phi_{\alpha(k)\dot\gamma(l-1)}{}^{\dot\beta} + h.c.
\end{equation}
Calculating variations of the term with the coefficient $\kappa_{k,l}$
we obtain (up to the terms that can be compensated by corrections to
graviton gauge transformations):
\begin{eqnarray}
\Delta_{k,l} &=& - R_{\dot\alpha\dot\beta} [ e_\beta{}^{\dot\gamma}
\Phi^{\alpha(k)\beta\dot\gamma(l-2)\dot\alpha} + a_{k,l}
e_{\beta\dot\delta} 
\Phi^{\alpha(k)\beta\dot\gamma(l-1)\dot\delta\dot\alpha} \nonumber \\
 && \qquad + b_{k,l} e^\alpha{}_{\dot\delta} 
\Phi^{\alpha(k-1)\dot\gamma(l-1)\dot\delta\dot\alpha} + c_{k,l}
e^{\alpha\dot\gamma} \Phi^{\alpha(k-1)\dot\gamma(l-2)\dot\alpha}]
 \zeta_{\alpha(k)\dot\gamma(l-1)}{}^{\dot\beta} \nonumber \\
 && + R_{\dot\alpha\dot\beta} 
\Phi_{\alpha(k)\dot\gamma(l-1)}{}^{\dot\alpha} [ 
  e_\beta{}^{\dot\gamma}
\zeta^{\alpha(k)\beta\dot\gamma(l-2)\dot\beta} + a_{k,l}
e_{\beta\dot\delta} 
\zeta^{\alpha(k)\beta\dot\gamma(l-1)\dot\delta\dot\beta} \nonumber \\
 && \qquad + b_{k,l} e^\alpha{}_{\dot\delta}
\zeta^{\alpha(k-1)\dot\gamma(l-1)\dot\delta\dot\beta} + c_{k,l}
e^{\alpha\dot\gamma} \zeta^{\alpha(k-1)\dot\gamma(l-2)\dot\beta}].
\end{eqnarray}
Now we consider gauge transformations with the parameters
$\zeta^{\alpha(k)\dot\alpha(l)}$ and collect all contributions. We
obtain
\begin{eqnarray}
\delta {\cal L}_1 &=& R_{\dot\alpha\dot\beta} [
((k+1)\kappa_{k+1,l-1}b_{k+1,l-1} - (l-1)\kappa_{k,l})
e_\beta{}^{\dot\delta} \Phi^{\alpha(k)\beta\dot\gamma(l-2)\dot\alpha} 
\zeta_{\alpha(k)\dot\gamma(l-2)\dot\delta}{}^{\dot\beta} \nonumber \\
 && \qquad - (\kappa_{k,l}a_{k,l} + (k+1)l\kappa_{k+1,l+1}c_{k+1,l+1})
e_{\beta\dot\delta}
\Phi^{\alpha(k)\beta\dot\gamma(l-1)\dot\delta\dot\alpha}
\zeta_{\alpha(k)\dot\gamma(l-1)}{}^{\dot\beta} \nonumber \\
 && \qquad + ( l\kappa_{k-1,l+1} - k\kappa_{k,l}b_{k,l} )
e^\beta{}_{\dot\delta} 
\Phi^{\alpha(k-1)\dot\gamma(l-1)\dot\delta\dot\alpha}
\zeta_{\alpha(k-1)\beta\dot\gamma(l-1)}{}^{\dot\beta} \nonumber \\
 && \qquad - ( k(l-1)\kappa_{k,l}c_{k,l} + \kappa_{k-1.l-1}a_{k-,l-1})
e^{\beta\dot\delta} \Phi^{\alpha(k-1)\dot\gamma(l-2)\dot\alpha}
\zeta_{\alpha(k-1)\beta\dot\gamma(l-2)\dot\delta}{}^{\dot\beta} ].
\end{eqnarray}
This gives us just two independent relations
\begin{equation}
l\kappa_{k-1,l+1} = kb_{k,l}\kappa_{k,l}, \qquad
a_{k-1,l-1}\kappa_{k-1,l-1} = - k(l-1)c_{k,l}\kappa_{k,l}.
\end{equation}
Here also we have to consider integer and half-integer spins
separately.

\subsection{Integer spin}

Switching on standard minimal gravitational interactions
\begin{equation}
e^{\alpha\dot\alpha} \Rightarrow e^{\alpha\dot\alpha} + g
h^{\alpha\dot\alpha}, \qquad D \Rightarrow D + g \omega^{\alpha(2)}
L_{\alpha(2)} + g \omega^{\dot\alpha(2)} L_{\dot\alpha(2)}
\end{equation}
breaks the gauge invariance of the initial Lagrangian (due to
non-commutativity of covariant derivatives) and leads to
\begin{eqnarray}
\delta \hat{\cal L}_0 &=& \sum_{m=1}^{s-1} (-1)^{(m+1)} 2g
[ \Omega_{\alpha(m)\beta\dot\alpha(m-1)} e^\beta{}_{\dot\beta}
( R^\alpha{}_\gamma \xi^{\alpha(m-1)\gamma\dot\alpha(m-1)\dot\beta}
+ R^{\dot\alpha}{}_{\dot\gamma}
\xi^{\alpha(m)\dot\alpha(m-2)\dot\beta\dot\gamma}) \nonumber \\
 && \qquad - f_{\alpha(m)\dot\alpha(m-1)\dot\beta}
e_\beta{}^{\dot\beta} ( R^\alpha{}_\gamma
\eta^{\alpha(m-1)\beta\gamma\dot\alpha(m-1)} 
+ R^{\dot\alpha}{}_{\dot\gamma}
\eta^{\alpha(m)\beta\dot\alpha(m-2)\dot\gamma}) ] + h.c. \label{non_1}
\end{eqnarray}
To compensate for this non-invariance we introduce non-minimal
interactions which, again using $k=m+n$, $l=m-n$, can be written as
follows:
\begin{equation}
{\cal L}_1 = \sum_{m=1}^{s-1} \sum_{n=0}^{m-1} \kappa_{m,n}
 R_{\dot\alpha\dot\beta} 
\Omega^{\alpha(m+n)\dot\gamma(m-n-1)\dot\alpha}
\Omega_{\alpha(m+n)\dot\gamma(m-n-1)}{}^{\dot\beta} + h.c. 
\end{equation}
Then the recurrent relations obtained in the previous subsection take
the form
\begin{eqnarray}
(m-n)\kappa_{m,n-1} &=& (m+n) b_{m,n}\kappa_{m,n}, \\
a_{m-1,n}\kappa_{m-1,n} &=& - (m+n)(m-n-1)c_{m,n}\kappa_{m,n}. 
\end{eqnarray}
Moreover, using explicit expressions for $a_{m,n}$ and $c_{m,n}$
$$
a_{m-1,n} = \frac{m(m-1)}{2(m-n)(m-n+1)}a_m, \quad m \ge 2,
$$
$$
c_{m,n} = \frac{1}{2(m+n)(m+n+1)}a_m,
$$
the second relation can be written as
\begin{equation}
\frac{m(m-1)}{(m-n)(m-n+1)}\kappa_{m-1,n}
= - \frac{(m-n-1)}{(m+n+1)}\kappa_{m,n}.
\end{equation}
Note that the first relation connects terms corresponding to the same
helicity but having different numbers of derivatives, while the second
relation connects terms with the same numbers of derivatives but
coming from different helicities.

Note that there are two ways to connect e.g. $\kappa_{m-1,n-1}$ and
$\kappa_{m,n}$ using these relations:
$$
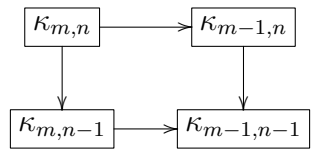
\xymatrix{
*+[F]{\kappa_{m,n}} \ar@{->}[d] \ar@{->}[r] &
*+[F]{\kappa_{m-1,n}} \ar@{->}[d] \\
*+[F]{\kappa_{m,n-1}} \ar@{->}[r] &
*+[F]{\kappa_{m-1,n-1}}  }
$$
From one hand we obtain
\begin{eqnarray*}
\kappa_{m-1,n-1} &=& \frac{(m+n-1)}{(m-n-1)}b_{m-1,n}\kappa_{m-1,n} \\
 &=& - \frac{(m+n-1)(m-n)(m-n+1)}{m(m-1)(m+n+1)}
b_{m-1,n}\kappa_{m,n},
\end{eqnarray*}
while from the other hand
\begin{eqnarray*}
\kappa_{m-1,n-1} &=& -
\frac{(m-n)(m-n+1)(m-n+2)}{m(m-1)(m+n)}\kappa_{m,n-1} \\
 &=& - \frac{(m-n+1)(m-n+2)}{m(m-1)} b_{m,n}\kappa_{m,n}. 
\end{eqnarray*}
But using a relation on parameters $b_{m,n}$
$$
b_{m,n} = \frac{(m+n-1)(m-n)}{(m+n+1)(m-n+2)}b_{m-1,n}
$$	
it is easy to check that the results coincide.

Schematically, the structure of non-minimal terms looks like:
$$
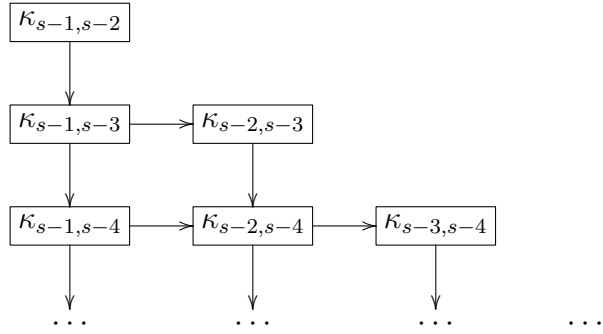
\xymatrix{
*+[F]{\kappa_{s-1,s-2}} \ar@{->}[d] &  &  & \\
*+[F]{\kappa_{s-1,s-3}} \ar@{->}[d] \ar@{->}[r] & 
*+[F]{\kappa_{s-2,s-3}} \ar@{->}[d] & & \\
*+[F]{\kappa_{s-1,s-4}} \ar@{->}[r] \ar@{->}[d] &
*+[F]{\kappa_{s-2,s-4}} \ar@{->}[r] \ar@{->}[d] &
*+[F]{\kappa_{s-3,s-4}} \ar@{->}[d] \\
\cdots & \cdots & \cdots & \cdots }
$$
Thus starting with the only term with the
highest number of derivatives and setting $\kappa_{s-1,s-2} = 1$, we
obtain  using the first relation
\begin{equation}
\kappa_{s-1,n} = \prod_{q=s-2}^{n+1} \frac{(s+q-1)}{(s-q-1)}
b_{s-1,q}.
\end{equation}
Let us stress that all dependencies on mass and the cosmological
constant enter through the parameters $b_{s-1,n}$, while the second
relation is purely combinatorial. This means that all terms having the
same number of derivatives have a common multiplier $\kappa_{s-1,n}$.

We still have to check what happens with the symmetries of the
Lagrangean fields. For
$\eta^{\alpha(m+1)\dot\alpha(m-1)}$-transformations we obtain
\begin{eqnarray}
\delta {\cal L}_1 &=& R_{\dot\alpha\dot\beta} [ 
a_{m+1}(\kappa_{m,1} + \frac{(m-1)}{(m+3)}\kappa_{m+1,1})
e^{\beta\dot\delta}
\Omega_{\alpha(m+1)\beta\dot\gamma(m-2)\dot\delta}{}^{\dot\alpha}
\eta^{\alpha(m+1)\dot\gamma(m-2)\dot\beta} \nonumber \\
 && \qquad + a_m(\frac{(m-2)}{(m+2)}\kappa_{m,1} + \kappa_{m-1,1})
e_{\beta\dot\delta} \Omega_{\alpha(m)\dot\gamma(m-3)}{}^{\dot\alpha}
\eta^{\alpha(m)\beta\dot\gamma(m-3)\dot\delta\dot\beta} \nonumber \\
 && \qquad ( - 2(m-2)\kappa_{m,1} + 2(m+2)b_{m,2}\kappa_{m,2})
e^\beta{}_{\dot\delta}
\Omega_{\alpha(m+1)\beta\dot\gamma(m-3)}{}^{\dot\alpha}
\eta^{\alpha(m+1)\dot\gamma(m-3)\dot\delta\dot\beta} \nonumber \\
 && \qquad +(b_m\kappa_{m,1} - 2(m-1)\kappa_{m,0}) 
f_{\alpha(m)\dot\gamma(m-2)\dot\delta}{}^{\dot\alpha}
e_\beta{}^{\dot\delta}
\eta^{\alpha(m)\beta\dot\gamma(m-2)\dot\beta} ] \nonumber \\
 && + 2\kappa_{m,0} R_{\alpha\beta} 
f^\alpha{}_{\gamma(m-1)\dot\alpha(m)} e_\delta{}^{\dot\alpha}
\eta^{\beta\gamma(m-1)\delta\dot\alpha(m-1)}.
\end{eqnarray}
With the help of relations
$$
\kappa_{m-1,1} = - \frac{(m-2)}{(m+2)}\kappa_{m,1}, \qquad
(m-2)\kappa_{m,1} = (m+2)b_{m,2}\kappa_{m,2},
$$
we finally get
\begin{eqnarray}
\delta {\cal L}_1 &=& - f_{\alpha(m)\dot\alpha(m-1)\dot\beta}
e_\beta{}^{\dot\beta} [2\kappa_{m,0} R^\alpha{}_\gamma
\eta^{\alpha(m-1)\beta\gamma\dot\alpha(m-1)} \nonumber \\
 && \qquad \qquad + (\frac{b_m}{(m-1)}\kappa_{m,1} - 2\kappa_{m,0})
 R^{\dot\alpha}{}_{\dot\gamma}
\eta^{\alpha(m)\beta\dot\alpha(m-2)\dot\gamma} ]. 
\end{eqnarray}
Similarly, for $\xi^{\alpha(m)\dot\alpha(m)}$-transformations we
obtain
\begin{eqnarray}
\delta {\cal L}_1 &=& R_{\dot\alpha\dot\beta} [
( \frac{m}{(m+2)}a_{m+1}(\kappa_{m,0} + \kappa_{m+1,0})
e^{\beta\dot\delta} 
f_{\alpha(m)\beta\dot\gamma(m-1)\dot\delta}{}^{\dot\alpha}
\xi^{\alpha(m)\dot\gamma(m-1)dot\beta} \nonumber \\
&& \qquad + \frac{(m-1)}{(m+1)}a_m (\kappa_{m,0} + \kappa_{m-1,0})
 e_{\beta\dot\delta} f_{\alpha(m-1)\dot\gamma(m-2)}{}^{\dot\alpha}
\xi^{\alpha(m-1)\beta\dot\gamma(m-2)\dot\delta\dot\beta} \nonumber \\
 && \qquad - ( - 2(m-1)\kappa_{m,0} + b_m\kappa_{m,1})
\Omega_{\alpha(m)\beta\dot\gamma(m-2)}{}^{\dot\alpha}
e^\beta{}_{\dot\delta}
\xi^{\alpha(m)\dot\gamma(m-2)\dot\delta\dot\beta} ] \nonumber \\
 && - 2\kappa_{m,0} R_{\alpha\beta} 
 \Omega^\alpha{}_{\gamma(m-1)\delta\dot\alpha(m-1)}
e^\delta{}_{\dot\alpha}
\xi^{\gamma(m-1)\beta\dot\alpha(m)}. 
\end{eqnarray}
Taking into account that
$$
\kappa_{m,0} = - \kappa_{m \pm 1,0},
$$
we get
\begin{eqnarray}
\delta {\cal L}_1 &=& \Omega_{\alpha(m)\beta\dot\alpha(m-1)}
e^\beta{}_{\dot\beta} [ 2\kappa_{m,0}
R^\alpha{}_\gamma
\xi^{\alpha(m-1)\gamma\dot\alpha(m-1)\dot\beta} \nonumber \\
 && \qquad \qquad + (\frac{b_m}{(m-1)}\kappa_{m,1} - 2\kappa_{m,0})
R^{\dot\alpha}{}_{\dot\gamma}
\xi^{\alpha(m)\dot\alpha(m-2)\dot\beta\dot\gamma} ].
\end{eqnarray}
Thus if we put
\begin{equation}
\kappa_{m,0} = \frac{b_m}{2(m-1)}\kappa_{m,1}, \qquad
g = (-1)^m\kappa_{m,0},
\end{equation}
then these variations cancel the non-invariance of the initial
Lagrangian (\ref{non_1}) due to non-commutativity of the covariant
derivatives. Note that it is easy to check that
$$
(-1)^m \frac{b_m}{2(m-1)}\kappa_{m,1}
= (-1)^{(m+1)}\frac{b_{m+1}}{2m}\kappa_{m+1,1},
$$
so that our solution is consistent with the relation $\kappa_{m,0} = -
\kappa_{m \pm 1,0}$. This in turn shows that the gravitational
coupling constant $g$ is universal and does not depend on helicity as
it should be. 

Let us use spin 4 case as an illustration.
\begin{eqnarray}
{\cal L}_1 &=& R_{\dot\alpha\dot\beta} [ \Omega^{\alpha(5)\dot\alpha}
\Omega_{\alpha(5)}{}^{\dot\beta} \nonumber \\
 && \qquad + \frac{(M^2-10\Lambda)}{10} ( 5 
\Omega^{\alpha(4)\dot\gamma\dot\alpha} 
\Omega_{\alpha(4)\dot\gamma}{}^{\dot\beta} -
\Omega^{\alpha(3)\dot\alpha} \Omega_{\alpha(3)}{}^{\dot\beta})
\nonumber \\
 && \qquad + \frac{(M^2-10\Lambda)(M^2-12\Lambda)}{12}
( f^{\alpha(3)\dot\gamma(2)\dot\alpha}
f_{\alpha(3)\dot\gamma(2)}{}^{\dot\beta}
- f^{\alpha(2)\dot\gamma\dot\alpha} 
f_{\alpha(2)\dot\gamma}{}^{\dot\beta} + f^{\alpha\dot\alpha}
f_\alpha{}^{\dot\beta}) ] \nonumber \\
 && + \frac{(M^2-10\Lambda)(M^2-12\Lambda)}{12} {\cal L}_{min}.
\end{eqnarray}
This concrete example shows the general properties of our solution.
Besides the massive case  in the unitary allowed region 
$M^2 > 12\Lambda$, there are only two cases where the standard minimal
gravitational interactions survive. The first one is well
known: the massless  limit in anti de Sitter space $\Lambda < 0$. The
second one is the partially massless limit at depth $k=2$ in de
Sitter space at $M^2 = 6\Lambda$. In all other partially massless
limits, minimal interactions (and even part of the non-minimal
ones) vanish.

\subsection{Half-integer spin}

In this case switching on minimal gravitational interaction leads to
\begin{equation}
\delta \hat{\cal L}_0 = \sum_{m=1}^{s-1} (-1)^m 
\Phi_{\alpha(m)\dot\alpha(m)\dot\beta} e_\beta{}^{\dot\beta}
( R^\alpha{}_\gamma \zeta^{\alpha(m-1)\beta\gamma\dot\alpha(m)} +
R^{\dot\alpha}{}_{\dot\gamma} 
\zeta^{\alpha(m)\beta\dot\alpha(m-1)\dot\gamma}). \label{non_2}
\end{equation}
Using $k=m+n+1$, $l=m-n$ an ansatz for non-minimal interactions can be
written as follows
\begin{equation}
{\cal L}_1 = R_{\dot\alpha\dot\beta} \sum_{m=1}^{s-1} \sum_{n=0}^{m-1}
\kappa_{m,n} \Phi^{\alpha(m+n+1)\dot\gamma(m-n-1)\dot\alpha}
\Phi_{\alpha(m+n+1)\dot\gamma(m-n-1)}{}^{\dot\beta} + h.c.
\end{equation} 
Then the recurrent relations take the form:
\begin{eqnarray}
(m-n)\kappa_{m,n-1}  &=& (m+n+1)b_{m,n}\kappa_{m,n}, \\
a_{m-1,n}\kappa_{m-1,n} &=& - (m-n-1)(m+n+1)c_{m,n}\kappa_{m,n}.
\end{eqnarray}
Moreover, using explicit expressions for $a_{m-1,n}$ and $c_{m,n}$
$$
a_{m-1,n} = \frac{m(m+1)}{(m-n)(m-n+1)}c_m,
$$
$$
c_{m,n} = \frac{(m+1)}{m(m+n+1)(m+n+2)}c_m, 
$$
the second relation takes the form
\begin{equation}
\frac{m}{(m-n)(m-n+1)}\kappa_{m-1,n} = 
- \frac{(m-n-1)}{m(m+n+2)}\kappa_{m,n}.
\end{equation}
As in the case of integer spin, the first relation connects terms
corresponding to the same helicity, but with different numbers of
derivatives, while the second relation connects terms with the same
number of derivatives, but coming from different helicities. Starting
with the only term with the highest number of derivatives
$\kappa_{s-1,s-2} = 1$, we can use the first relation to obtain
\begin{equation}
\kappa_{s-1,n} = \prod_{q=s-2}^{n+1} \frac{(s+q)}{(s-q-1)}b_{s-1,q},
\qquad n < s-2.
\end{equation}

We still have to check what happens with the symmetries of the
Lagrangean fields. For
$\zeta^{\alpha(m+1)\dot\alpha(m)}$-transformations we obtain
\begin{eqnarray}
\delta {\cal L}_1 &=& R_{\dot\alpha\dot\beta} [
2[ (m-1)\kappa_{m,0} - (m+2)b_{m,1}\kappa_{m,1}]
e^\beta{}_{\dot\delta} 
\Phi_{\alpha(m+1)\beta\dot\gamma(m-2)}{}^{\dot\alpha}
\zeta^{\alpha(m+1)\dot\gamma(m-2)\dot\delta\dot\beta} \nonumber \\
 && \qquad - 2c_{m+1}[ \kappa_{m,0} +
\frac{m(m+2)}{(m+1)(m+3)}\kappa_{m+1,0}] e^{\beta\dot\delta} 
\Phi_{\alpha(m+1)\beta\dot\gamma(m-1)\dot\delta}{}^{\dot\alpha}
 \zeta^{\alpha(m+1)\dot\gamma(m-1)\dot\beta} \nonumber \\
 && \qquad - 2c_m[ \frac{(m+1)(m-1)}{m(m+2)}\kappa_{m,0} 
+ \kappa_{m-1,0}]
 e_{\beta\dot\beta} \Phi_{\alpha(m)\dot\gamma(m-2)}{}^{\dot\alpha}
 \zeta^{\alpha(m)\beta\dot\gamma(m-2)\dot\beta} \nonumber  \\
 && \qquad + 2(m+1)d_m\kappa_{m,0} e_\beta{}^{\dot\delta}
\Phi_{\alpha(m)\dot\gamma(m-1)\dot\delta}{}^{\dot\alpha}
\zeta^{\alpha(m)\beta\dot\gamma(m-1)\dot\beta} ] \nonumber \\
 && - 2d_m\kappa_{m,0} R_{\alpha\beta}
\Phi^\alpha{}_{\gamma(m-1)\dot\gamma(m+1)}
e_\delta{}^{\dot\gamma}
\zeta^{\beta\delta\gamma(m-1)\dot\gamma(m)}.
\end{eqnarray}
First of all we put
\begin{equation}
\kappa_{m,0} = \frac{(m+2)}{(m-1)}b_{m,1}\kappa_{m,1}.
\end{equation}
It is easy to check that this solution is consistent with the
relation
$$
\kappa_{m,0} = \frac{m(m+2)}{(m+1)(m+3)}\kappa_{m+1,0}.
$$
Thus we finally get
\begin{equation}
\delta {\cal L}_1 = \frac{2(m+1)}{m}d_m\kappa_{m,0}
\Phi_{\alpha(m)\dot\alpha(m)\dot\beta} e_\beta{}^{\dot\beta}
[R^\alpha{}_\gamma \zeta^{\alpha(m-1)\beta\gamma\dot\alpha(m)} 
+ R^{\dot\alpha}{}_{\dot\gamma}
\zeta^{\alpha(m)\beta\dot\alpha(m-1)\dot\gamma}  ]. 
\end{equation} 
Thus if we put
\begin{equation}
g = (-1)^m\frac{(m+1)}{m}d_m\kappa_{m,0} 
\end{equation}
these variations cancel the non-invariance of the initial Lagrangian
(\ref{non_2}). Here also it can be easily checked that
$$
(-1)^m\frac{(m+1)}{m}d_m\kappa_{m,0}
= (-1)^{(m+1)}\frac{(m+2)}{(m+1)}d_{m+1}\kappa_{m+1,0},
$$
so that gravitational coupling constant does not depend on helicity.

Let us use a spin 9/2 case as an example	.
\begin{eqnarray}
{\cal L}_1 &=& R_{\dot\alpha\dot\beta} [ \Phi^{\alpha(6)\dot\alpha}
\Phi_{\alpha(6)}{}^{\dot\beta} \nonumber \\
 && \quad + \frac{(M^2-12\Lambda)}{2} 
( \Phi^{\alpha(5)\dot\gamma\dot\alpha}
\Phi_{\alpha(5)\dot\gamma}{}^{\dot\beta} - \frac{1}{9}
\Phi^{\alpha(4)\dot\alpha} \Phi_{\alpha(4)}{}^{\dot\beta}) \nonumber 
\\
 && \quad + \frac{(M^2-12\Lambda)(M^2-15\Lambda)}{12}
( \Phi^{\alpha(4)\dot\gamma(2)\dot\alpha}
\Phi_{\alpha(4)\dot\gamma(2)}{}^{\dot\beta}
- \frac{2}{3} \Phi^{\alpha(3)\dot\gamma\dot\alpha}
\Phi_{\alpha(3)\dot\gamma}{}^{\dot\beta} + \frac{1}{4}
\Phi^{\alpha(2)\dot\alpha} \Phi_{\alpha(2)}{}^{\dot\beta}) ] \nonumber
\\
 && + \frac{(M^2-12\Lambda)(M^2-15\Lambda)\sqrt{M^2-16\Lambda}}{36}
{\cal L}_{min}.
\end{eqnarray}
These results show the general properties of our solution, namely,
besides the massless fields in anti de Sitter space, only partially
massless fields with a depth $t=2$ ($M^2 = 7\Lambda$ for spin 9/2)
have standard minimal gravitational interactions. For all other
partially massless fields, as well as for fields living on the
boundary of the unitary forbidden region, minimal interactions (and
part of the non-minimal interactions) vanish.

\section{Conclusion}

In this work, we constructed gravitational vertices for massive fields
with arbitrary integer and half-integer spins, which contain
standard minimal interactions as well as non-minimal interactions
necessary to make the vertex gauge invariant. The main assumption was
that there exists a non-singular massless limit in anti de Sitter
space, meaning that the  highest number of derivatives must be the
equal to that in the massless case. An ansatz for appropriate
non-minimal terms was proposed, and it was shown that it leads to a
unique solution that correctly reproduces our previous results
including all possible partially massless limits.

\end{document}